\documentclass[12pt]{article}
\usepackage{latexsym}
\usepackage{amsmath,amsfonts}
\usepackage{times}
\allowdisplaybreaks[4]

\catcode`\@=11

\newcount\hour
\newcount\minute
\newtoks\amorpm \hour=\time\divide\hour by 60\minute
=\time{\multiply\hour by 60 \global\advance\minute by-\hour}
\edef\standardtime{{\ifnum\hour<12 \global\amorpm={am}%
        \else\global\amorpm={pm}\advance\hour by-12 \fi
        \ifnum\hour=0 \hour=12 \fi
        \number\hour:\ifnum\minute<10
        0\fi\number\minute\the\amorpm}}
\edef\militarytime{\number\hour:\ifnum\minute<10 0\fi\number\minute}

\def\draftlabel#1{{\@bsphack\if@filesw {\let\thepage\relax
   \xdef\@gtempa{\write\@auxout{\string
      \newlabel{#1}{{\@currentlabel}{\thepage}}}}}\@gtempa
   \if@nobreak \ifvmode\nobreak\fi\fi\fi\@esphack}
        \gdef\@eqnlabel{#1}}
\def\@eqnlabel{}
\def\@vacuum{}
\def\marginnote#1{}
\def\draftmarginnote#1{\marginpar{\raggedright\scriptsize\tt#1}}
\def\draft{
        \pagestyle{plain}
        \overfullrule=2pt
        \oddsidemargin -.5truein
        \def\@oddhead{\sl \phantom{\today\quad\militarytime} \hfil
        \smash{\Large\sl DRAFT} \hfil \today\quad\militarytime}
        \let\@evenhead\@oddhead
        \let\label=\draftlabel
        \let\marginnote=\draftmarginnote
        \def\ps@empty{\let\@mkboth\@gobbletwo
        \def\@oddfoot{\hfil \smash{\Large\sl DRAFT} \hfil}
        \let\@evenfoot\@oddhead}
        \def\@eqnnum{(\theequation)\rlap{\kern\marginparsep\tt\@eqnlabel}%
        \global\let\@eqnlabel\@vacuum}  }

\newcommand{\rf}[1]{(\ref{#1})}
\renewcommand{\theequation}{\thesection.\arabic{equation}}
\renewcommand{\thefootnote}{\fnsymbol{footnote}}
\newcommand{\newsection}{   
\setcounter{equation}{0}\section}

\def\appendix#1{\addtocounter{section}{1}\setcounter{equation}{0}
\renewcommand{\thesection}{\Alph{section}}
\section*{Appendix \thesection\protect\indent \parbox[t]{11.15cm}{#1}}
\addcontentsline{toc}{section}{Appendix \thesection\ \ \ #1}}

\def\be{\begin{equation}}
\def\ee{\end{equation}}
\def\beq{\begin{eqnarray}}
\def\eeq{\end{eqnarray}}

\def\parline{\,\partial\kern -0.55em /\,\,}

\def\half{{\frac{1}{2}}}

\def\DD{{\cal D}}

\def\LL{{\cal L}}

\def\ZZ{{\cal Z}}

\def\ibf{{\bf i}}
\def\iibf{{\bf ii}}
\def\iiibf{{\bf iii}}
\def\ivbf{{\bf iv}}
\def\vbf{{\bf v}}

\def\ck{|c\rangle}
\def\cbk{|\bar{c}\rangle}
\def\bk{|b\rangle}

\def\phik{|\phi\rangle}
\def\phibr{\langle\phi|}
\def\bbr{\langle b |}

\def\xik{|\xi\rangle}

\def\smponetwo{{\scriptscriptstyle [1,2]}}

\def\sm(A)dS{{\scriptscriptstyle (A)dS }}

\def\Lb{\bar{L}}
\def\cb{\bar{c}}

\def\alphab{\bar\alpha}
\def\upsilonb{\bar\upsilon}

\def\eb{\bar{e}}

\def\I{{\rm I}}
\def\II{{\rm II}}

\def\tot{{\rm tot}}
\def\qu{{\rm qu}}
\def\FP{{\rm FP}}

\def\mun{{\underline{m}}}

\def\ssf{{\sf s}}
\def\ssfb{\bar{\sf s}}

\begin{document}


\begin{flushright}
FIAN-TD-2016-22  \hspace{1.5cm} \ \ \ \ \  \\
arXiv: 1610.00657 V2 [hep-th] \\
\end{flushright}

\vspace{1cm}

\begin{center}

{\Large \bf Continuous spin gauge field in (A)dS space}

\vspace{2.5cm}

R.R. Metsaev%
\footnote{ E-mail: metsaev@lpi.ru
}

\vspace{1cm}

{\it Department of Theoretical Physics, P.N. Lebedev Physical
Institute, \\ Leninsky prospect 53,  Moscow 119991, Russia }

\vspace{3.5cm}

{\bf Abstract}

\end{center}

Totally symmetric continuous spin field propagating in (A)dS is studied.
Lagrangian gauge invariant formulation for such field is developed.
Lagrangian of continuous spin field is constructed in terms of double traceless tensor fields, while gauge
transformations are constructed in terms of traceless gauge transformation parameters.
de Donder like gauge condition that leads to simple gauge fixed Lagrangian is found.
Gauge-fixed Lagrangian invariant under global BRST transformations is presented.
The BRST Lagrangian is used for computation of a partition function. It is demonstrated that the partition
function of the continuous spin field is equal to one.
Various decoupling limits of the continuous spin field are also studied.

\vspace{2cm}

Keywords: continuous spin field; higher-spin field.

\newpage
\renewcommand{\thefootnote}{\arabic{footnote}}
\setcounter{footnote}{0}

\newsection{\large Introduction}

Continuous spin field has attracted some interest in recent time.  Such field can be considered as
a field theoretical realization of continuous spin representation of Poincar\'e algebra which was studied many years ago in Ref.\cite{Bargmann:1948ck}. For extensive list of references on this theme see Refs.\cite{Brink:2002zx,Bekaert:2005in}. Interesting feature of continuous spin field is that
this field is decomposed into infinite chain of coupled scalar, vector, and tensor fields which consists of every field just once. We note then that a similar infinite chain of fields enters higher-spin gauge field theories in AdS space \cite{Vasiliev:1990en}. Note however that fields in Ref.\cite{Vasiliev:1990en} are decoupled as coupling constant tends to zero.  Also it turns out that some regimes in string theory
are related to continuous spin field \cite{Savvidy:2003fx}.
We think that further progress in understanding dynamics of continuous spin field requires, among
other things, better understanding of gauge invariant Lagrangian formulation of continuous spin field in (A)dS and flat spaces. This is what we are doing in this paper.

Gauge invariant formulation for bosonic continuous spin field in four-dimensional flat space, $R^{3,1}$, was developed in Ref.\cite{Schuster:2014hca}, while gauge theory of fermionic continuous spin field in $R^{3,1}$ was studied in Ref.\cite{Najafizadeh:2015uxa}. So far Lagrangian formulation of continuous spin field propagating in (A)dS space has not been discussed in the literature. Our major aim in this paper is to develop Lagrangian gauge invariant formulation of continuous spin bosonic field in $(A)dS_{d+1}$ space with arbitrary $d\geq 3$. We use our gauge invariant Lagrangian for derivation of gauge-fixed BRST Lagrangian of continuous spin field which is invariant under global BRST and anti-BRST transformations. We use our BRST Lagrangian for computation of a partition function and demonstrate that such partition function is equal to 1. Also we analyse various limits of gauge invariant Lagrangian for continuous spin field in (A)dS space. We demonstrate that such limits lead to appearance of massless, massive and partial-massless fields. By product, considering limit of flat space, we obtain Lagrangian gauge invariant formulation of continuous spin field in flat $R^{d,1}$ with arbitrary $d\geq 3$.  We note that, so far, Lagrangian formulation of continuous spin field in flat space $R^{d,1}$ with arbitrary $d\geq 3$ was discussed only in the framework of light-cone gauge approach \cite{Brink:2002zx}.

\newsection{ \large Lagrangian and gauge transformations of continuous spin field}

We start with a discussion of a field content entering our gauge invariant
formulation of continuous spin field. To discuss a continuous spin
field propagating in $AdS_{d+1}$ space, we introduce scalar, vector and tensor fields of the $so(d,1)$ Lorentz algebra,

\be \label{man-19092016-01}
\phi^{a_1\ldots a_n}\,, \hspace{2cm} n=0,1,\ldots,\infty\,.
\ee
In \rf{man-19092016-01}, fields with $n=0$ and $n=1$ are the respective scalar and vector fields of the $so(d,1)$ algebra, while fields with $n \geq 2$ are the totally symmetric tensor fields of the Lorentz  $so(d,1)$ algebra.  Fields $\phi^{a_1\ldots a_n}$ \rf{man-19092016-01} with $n \geq 4$
are taken to be double-traceless,

\be  \label{man-19092016-02}
\phi^{aabba_5\ldots a_n}=0\,, \hspace{2cm}
n=4,5,\ldots,\infty.
\ee
Fields in \rf{man-19092016-01} subject to constraint \rf{man-19092016-02}
constitute a field content of our approach.

To streamline our presentation we introduce a set of creation operators $\alpha^a$, $\upsilon$,
and the respective set of annihilation operators, $\alphab^a$,
$\upsilonb$ which we will refer to as oscillators. Using the $\alpha^a$, $\upsilon$, we collect fields \rf{man-19092016-01}
into a ket-vector $|\phi\rangle$ defined as

\be \label{man-20092016-01}
\phik = \sum_{n=0}^\infty \frac{\upsilon^n}{n!\sqrt{n!}} \alpha^{a_1} \ldots \alpha^{a_n} \phi^{a_1\ldots a_n} |0\rangle\,.
\ee
In terms of the ket-vector $\phik$, constraint \rf{man-19092016-02}
can be represented as $(\alphab^2)^2 \phik = 0$.

Gauge invariant action and Lagrangian of continuous spin field we found can be presented as
\beq
\label{man-20092016-04} S  & = &  \int d^{d+1}x\,\LL\,,  \qquad  \LL  = \half e \phibr   E \phik \,,
\\
\label{man-20092016-05} E & \equiv & (1-\frac{1}{4}\alpha^2 \bar\alpha^2) (\Box_\sm(A)dS + m_1 + m_2
\alpha^2\bar\alpha^2) - L \Lb\,,
\\
\label{man-20092016-06} && \Lb \equiv  \bar\alpha D - \half \alpha D  \bar\alpha^2  -
\eb_1\Pi^\smponetwo + \half e_1 \bar\alpha^2\,,
\\
\label{man-20092016-07} && L \equiv \alpha D  - \half \alpha^2 \bar\alpha D  - e_1 \Pi^\smponetwo +
\half \eb_1 \alpha^2\,,
\eeq
$\langle\phi| \equiv (\phik)^\dagger$, where $e=\det e_\mun^a$, while $e_\mun^a$ stands for vielbein in (A)dS space. The notation  $\Box_{(A)dS}$ in \rf{man-20092016-05} is used for the  D'Alembert operator in (A)dS space. Quantities $m_1$, $m_2$, and $e_1$, $\eb_1$ appearing in  \rf{man-20092016-05}-\rf{man-20092016-07} are defined by relations
\beq
\label{man-20092016-08} && m_1  = - \mu_0 - \rho \Bigl( N_\upsilon(N_\upsilon+d-1) + 2d - 4 \Bigr) \,, \qquad m_2  =  \rho\,,
\\
\label{man-20092016-10} && e_1 =  e_\upsilon \upsilonb\,, \qquad  \eb_1 = - \upsilon e_\upsilon\,,
\\
\label{man-20092016-12} && e_\upsilon  = \Bigl[\frac{1}{(N_\upsilon + 1) (2N_\upsilon+d-1)} F(N_\upsilon)\Bigr]^{1/2} \,,
\\
\label{man-20092016-14} && F( N_\upsilon) \equiv   \mu_1 -  N_\upsilon (N_\upsilon + d-2) \Bigl(\mu_0 + \rho (N_\upsilon + 1) (N_\upsilon +d-3)  \Bigr) \,,
\eeq
where, in \rf{man-20092016-08},\rf{man-20092016-14},  $\mu_0$, $\mu_1$ stand for dimensionfull constants, while $\rho$ is defined as

\be \label{man-20092016-15}
\rho = -\frac{1}{R^2} \ \hbox{ for AdS space}; \quad \rho = 0 \ \hbox{ for flat space}; \quad \rho = \frac{1}{R^2} \ \hbox{ for dS space},
\ee
and $R$ is a radius of (A)dS space. Quantities $N_\upsilon$, $\alpha D$, $\alpha^2$ are defined in Appendix. We note the relation $e\phibr L\Lb\phik = - e \langle \Lb\phi|\Lb\phik$ (up to total derivative). The following remarks are in order.

\noindent \ibf) Our Lagrangian depends on $\rho$ given in \rf{man-20092016-15} and two arbitrary dimensionfull parameters $\mu_0$, $\mu_1$.

\noindent \iibf) On space of double-traceless ket-vector $\phik$, operator $E$ \rf{man-20092016-05} can alternatively be represented as

\beq
\label{man-20092016-16} E & = & \Box_\sm(A)dS + M_1 -\frac{1}{4}\alpha^2\bar\alpha^2 (\Box_\sm(A)dS + M_2)  - L \Lb\,,
\\
\label{man-20092016-17} && M_1  \equiv  - \mu_0 - \rho \bigl( N_\upsilon(N_\upsilon+d-1) + 2d - 4 \bigr),
\\
\label{man-20092016-18} && M_2 \equiv  - \mu_0 - \rho \bigl( N_\upsilon(N_\upsilon+d-5) + 6  \bigr).
\eeq

\noindent \iiibf) Two-derivative contributions to operator $E$ \rf{man-20092016-05} coincide with two-derivative contributions to the standard Fronsdal operator that enters Lagrangian
of free massless field in (A)dS space.

\noindent \ivbf) Representation for gauge invariant Lagrangian given in \rf{man-20092016-04}-\rf{man-20092016-07} is
universal and is valid for arbitrary theory of gauge fields propagating in (A)dS space. Various
(A)dS field theories are distinguished by operators $m_1$, $m_2$, $e_1$, $\eb_1$ entering
the operator $E$. Namely, the operators $E$ of massless, massive,
conformal, and continuous spin fields propagating in (A)dS space depend on the covariant derivative $D^a$ and the oscillators $\alpha^a$, $\bar\alpha^a$  in the same way as the
operator $E$ given in \rf{man-20092016-04}. This is to say that
operators $E$ for massless, massive, conformal, and continuous spin field in (A)dS space are distinguished only by the operators $m_1$, $m_2$, $e_1$, and $\eb_1$. It is finding the operators $m_1$, $m_2$, $e_1$, and $\eb_1$ that provides real difficulty. For the reader convenience we note that, for massless fields in $(A)dS_{d+1}$, the operators $m_1$, $m_2$, $e_1$, and $\eb_1$ take the form

\beq
\label{man-20092016-19} && m_1 = \rho \bigl( N_\upsilon(N_\upsilon+d-5) -2d+4 \bigr), \qquad m_2  =  \rho,
\nonumber\\
\label{man-20092016-20} && e_1  = 0, \qquad \eb_1= 0, \qquad \hbox{ for massless fields in } (A)dS_{d+1}\,.
\eeq
Explicit expressions for the operators $m_1$, $m_2$, $e_1$, and $\eb_1$ corresponding to the massive and conformal fields in (A)dS can be found in Refs.\cite{Metsaev:2009hp,Metsaev:2014iwa}.

\noindent \vbf) It is the use of operators $L$, $\Lb$ \rf{man-20092016-06},\rf{man-20092016-07} that considerably simplifies our Lagrangian. We refer to the operator $\Lb$ as modified de Donder divergence. Equating $e_1=0$, $\eb_1=0$ gives the standard de Donder divergence. For massless continuous spin field in $R^{3,1}$, i.e., the case $d=3$, $\mu_0=0$, $\rho=0$, the operator $\Lb$ was introduced in Ref.\cite{Schuster:2014hca}. For $d=3$, $\mu_0=0$, $\rho=0$, our Lagrangian \rf{man-20092016-04} coincides with the one in Ref.\cite{Schuster:2014hca}. Idea to use modified de Donder divergence to simplify Lagrangian of massive field in flat and (A)dS space was first exploited in Refs\cite{Metsaev:2008fs,Metsaev:2014iwa}. Alternative representation for Lagrangian of massive field without use of de Donder was first obtained in Ref.\cite{Zinoviev:2001dt}. Discussion of the standard de Donder divergence for studying various aspects of higher-spin field theory may be found in Refs.\cite{Francia:2007qt}.

{\bf Gauge symmetries}. To discuss gauge symmetries of continuous spin field we introduce
the following gauge transformation parameters:

\be \label{man-21092016-01}
\xi^{a_1\ldots a_n}\,,\qquad\qquad n=0,1,\ldots, \infty\,.
\ee
In \rf{man-21092016-01}, gauge parameters with $n=0$ and $n=1$ are the
respective scalar and vector fields of the Lorentz $so(d,1)$ algebra, while the gauge parameters with $n\geq 2$ are totally symmetric traceless tensor fields of the Lorentz $so(d,1)$ algebra,

\be \label{man-21092016-02}
\xi^{aaa_3\ldots a_n}=0\,, \qquad n= 2,3,\ldots, \infty\,.
\ee
To streamline presentation of gauge symmetries we use the $\alpha^a$, $\upsilon$ and collect gauge transformation parameters in ket-vector $\xik$ defined as

\be \label{man-21092016-03}
\xik = \sum_{n=0}^\infty \frac{\upsilon^{n+1}}{n!\sqrt{(n+1)!}} \alpha^{a_1} \ldots \alpha^{a_n} \xi^{a_1\ldots a_n} |0\rangle\,.
\ee
Note also that, in terms of the $\xik$,  algebraic constraints \rf{man-21092016-02} take the form $\alphab^2 \xik=0$.

Using ket-vectors $\phik$ and $\xik$ we note that Lagrangian for continuous spin field \rf{man-20092016-04} is invariant under the following gauge transformations:

\be \label{man-21092016-06}
\delta \phik  = G\xik\,, \qquad G = \alpha D - e_1 - \alpha^2\frac{1}{2N_\alpha
+ d- 1}\eb_1 \,,
\ee
where the operators $e_1$, $\eb_1$ appearing in \rf{man-21092016-06} are defined in \rf{man-20092016-10}-\rf{man-20092016-14}.

\newsection{\large  BRST Lagrangian and partition function of continuous spin field}\label{secti-03}

{\bf BRST invariant Lagrangian of continuous spin field}. In this section we obtain gauge-fixed BRST invariant Lagrangian for continuous spin field. We use then such Lagrangian to compute a partition function of the continuous spin field. As we have already said a general structure of our Lagrangian \rf{man-20092016-04} and gauge transformations \rf{man-21092016-06} for continuous spin (A)dS field is similar to the one for massive (A)dS field. Derivation of BRST Lagrangian and use of such Lagrangian for a computation of partition function of massive (A)dS field may be found in Ref.\cite{Metsaev:2014vda}. In this section we demonstrate how the method in Ref.\cite{Metsaev:2014vda} can be applied to the case of continuous spin field. Lagrangian of continuous spin field with local BRST symmetries was discussed in Ref.\cite{Bengtsson:2013vra}.

To built gauge-fixed BRST invariant Lagrangian we introduce Faddeev-Popov fields $c^{a_1\ldots a_n}$, $\cb^{a_1\ldots a_n}$ and Nakanishi-Lautrup fields $b^{a_1\ldots a_n}$, $n=0,1,\ldots, \infty$.
We use the oscillators
to collect all Faddeev-Popov fields into  ket-vectors $\ck$, $\cbk$, while all Nakanishi-Lautrup fields are collected into ket-vector $\bk$. Using notation $|\chi\rangle$ for $\ck$, $\cbk$, $\bk$, we note that  representation of the ket-vectors $\ck$, $\cbk$, $\bk$ in terms of scalar, vector, and tensor fields of  the $so(d,1)$ algebra takes the form
\be
\label{man-22092016-01}  |\chi\rangle = \sum_{n=0}^\infty \frac{\upsilon^{n+1}}{n!\sqrt{(n+1)!}} \alpha^{a_1} \ldots \alpha^{a_n} \chi^{a_1\ldots a_n} |0\rangle\,,\qquad \chi = (c,\cb,b)\,.
\ee
Fields in \rf{man-22092016-01} with $n=0$, $n=1$, and $n\geq 2$ are the respective scalar, vector, and traceless totally symmetric tensor fields of the $so(d,1)$ algebra.

Using ket-vector $\phik$ \rf{man-20092016-01} and ket-vectors in \rf{man-22092016-01}, we note that gauge-fixed Lagrangian $\LL_\tot$ in arbitrary  $\alpha$-gauge can be presented as
\beq
\label{man-22092016-04} && \LL_\tot = \LL + \LL_\qu \,, \qquad e^{-1} \LL_\qu = - \langle b |\Lb\phik + \langle \cb| ( \Box_\sm(A)dS + M_{_\FP} \bigr) \ck + \half \alpha \bbr\bk\,,\qquad
\\
\label{man-22092016-06} && \hspace{4cm} M_{_\FP}   =   -\mu_0 - \rho \Bigl( N_\upsilon(N_\upsilon+d-3) +  d - 2  \Bigr) \,,
\eeq
where gauge invariant Lagrangian $\LL$ is given in \rf{man-20092016-04}, while the modified de Donder operator $\Lb$ is defined in \rf{man-20092016-06}. One can verify that, up to total derivative, gauge-fixed Lagrangian \rf{man-22092016-04} is invariant under the following BRST and anti-BRST transformations
\beq
\label{man-22092016-07} &&  \ssf  \phik =   G \ck\,,  \qquad \ssf  \ck  =  0  \,,  \hspace{1.6cm} \ssf  |\cb\rangle  =    \bk \,, \qquad \ssf   \bk = 0 \,,\qquad
\\
\label{man-22092016-08} && \ssfb   \phik =   G |\cb\rangle\,,\qquad \ssfb   \ck  = -  \bk \,,\hspace{1cm}  \ssfb   |\cb\rangle  = 0\,, \hspace{1.2cm} \ssfb   \bk  =  0  \,,
\eeq
where gauge transformation operator $G$ is defined in \rf{man-21092016-06}. It easy to check then that BRST and anti-BRST transformations given in \rf{man-22092016-07}, \rf{man-22092016-08} are off-shell nilpotent: $\ssf^2=0$, $\ssfb^2=0$, $\ssf\ssfb+\ssfb\ssf=0$.

Lagrangian \rf{man-22092016-04} can be cast into the form that is more convenient for practical calculations. This is to say that fixing  the $\alpha=1$ gauge and integrating out Nakanishi-Lautrup fields, we find that Lagrangian \rf{man-22092016-04} leads to the following gauge-fixed Lagrangian
\be \label{man-22092016-09}
e^{-1} \LL_\tot = \half \phibr (1-\frac{1}{4}\alpha^2\alphab^2) \bigl( \Box_\sm(A)dS + m_1 +m_2 \alpha^2 \alphab^2 \bigr)\phik  + \langle \cb| ( \Box_\sm(A)dS + M_{_\FP} \bigr) \ck\,.\qquad
\ee
Alternatively, Lagrangian \rf{man-22092016-09} can be represented in terms of $M_1$, $M_2$ \rf{man-20092016-17},\rf{man-20092016-18} as
\be \label{man-23092016-01}
e^{-1} \LL_\tot =  \half  \phibr\Bigl( \Box_\sm(A)dS + M_1 -  \frac{1}{4}\alpha^2 \alphab^2 ( \Box_\sm(A)dS + M_2) \Bigr)\phik +  \langle \cb| ( \Box_\sm(A)dS + M_{_\FP} \bigr) \ck\,,
\ee
where $M_\FP$ is given in \rf{man-22092016-06}.  Thus we see that it is the use of representation for gauge invariant Lagrangian in \rf{man-20092016-05}-\rf{man-20092016-07} and the $\alpha=1$ gauge that simplify considerably the expression for gauge-fixed Lagrangian given in \rf{man-22092016-09},\rf{man-23092016-01}. Gauge-fixed Lagrangian given in \rf{man-22092016-09} and \rf{man-23092016-01} is also invariant under BRST and anti-BRST transformations given by

\be
\label{man-23092016-02} \ssf \phik
=
G  \ck\,, \quad
\ssf  \ck = 0\,, \quad
\ssf \cbk
=  \Lb
\phik \,, \quad \ssfb  \phik  =
G  \cbk\,,
\quad
\ssfb  \ck =
-
\Lb \phik\,,
\quad \ssfb   \cbk   = 0\,,
\ee
where $\Lb$ and $G$ are given in \rf{man-20092016-06} and \rf{man-21092016-06}.
Note however that, in contrast to transformations given in \rf{man-22092016-07},\rf{man-22092016-08}, BRST and anti-BRST transformations given in \rf{man-23092016-02} are nilpotent, $\ssf^2=0$, $\ssfb^2=0$, $\ssf\ssfb+\ssfb\ssf=0$, only for on-shell Faddeev-Popov fields.

\noindent {\bf Partition function of continuous spin field}. In order to compute a partition function of continuous spin field we decompose double-traceless ket-vector $\phik$ \rf{man-20092016-01} into two traceless ket-vectors denoted by $|\phi_{_\I} \rangle$, $|\phi_{_\II} \rangle$,
\be \label{man-23092016-04}
\phik = |\phi_{_\I} \rangle   + \alpha^2 \bigl((2N_\alpha+d-1)(2N_\alpha+d+1) \bigr)^{-1/2} |\phi_{_\II} \rangle, \qquad  \alphab^2 |\phi_{_\I} \rangle =0,  \quad \alphab^2 |\phi_{_\II} \rangle =0.
\ee
Using relations \rf{man-20092016-01},\rf{man-23092016-04}, it easy to understand that a decomposition of the traceless ket-vectors $|\phi_{_\I} \rangle$,  $|\phi_{_\II} \rangle$ into scalar, vector, and traceless tensor fields of the $so(d,1)$ algebra can be presented as
\beq
\label{man-24092016-01} && |\phi_{_\I}\rangle = \sum_{n=0}^\infty \frac{\upsilon^n}{n!\sqrt{n!}} \alpha^{a_1} \ldots \alpha^{a_n} \phi_{_\I}^{a_1\ldots a_n} |0\rangle\,,
\\
\label{man-24092016-02} && |\phi_{_\II}\rangle = \sum_{n=0}^\infty \frac{\upsilon^{n+2}}{n!\sqrt{(n+2)!}} \alpha^{a_1} \ldots \alpha^{a_n} \phi_{_\II}^{a_1\ldots a_n} |0\rangle\,.
\eeq
Plugging $\phik$ \rf{man-23092016-04} into gauge-fixed Lagrangian \rf{man-23092016-01}, we get
\beq
\label{man-24092016-03} && \hspace{-1cm} \LL_\tot =    \LL_\I  -  \LL_\II  +   \LL_{_\FP}\,,
\\
\label{man-24092016-04} && \LL_\tau  \equiv \frac{e}{2} \langle \phi_\tau| \bigl( \Box_\sm(A)dS + M_\tau \bigr) |\phi_\tau\rangle\,,\qquad \tau = \I,\II
\\
\label{man-24092016-06} && \LL_{_\FP} \equiv   e \langle \cb| \bigl( \Box_\sm(A)dS +  M_{_\FP} \bigr) |c \rangle\,,
\\
\label{man-24092016-07} && \hspace{1cm} M_{_\I}   =    -\mu_0 - \rho \Bigl( N_\upsilon(N_\upsilon+d-1) + 2d - 4 \Bigr) \,,
\\
\label{man-24092016-08} && \hspace{1cm} M_{_\II}  =    -\mu_0 - \rho \Bigl( N_\upsilon(N_\upsilon+d-5) + 2  \Bigr) \,,
\eeq
where operator $M_{_\FP}$ is defined in \rf{man-22092016-06}. We note that operators $M_{_\I}$, $M_{_\II}$ \rf{man-24092016-07}, \rf{man-24092016-08} are related to operators $M_1$, $M_2$ \rf{man-20092016-17},\rf{man-20092016-18} as $M_{_\I} = M_1$, $M_{_\II} = M_2 + 4\rho$.
In terms of scalar, vector, and tensor fields expressions given in \rf{man-24092016-04},\rf{man-24092016-06} can be represented as
\beq
\label{man-24092016-10} && \hspace{-2cm} \LL_{_\I} \equiv  \sum_{n=0}^\infty \LL_{_\I}^n\,, \qquad \LL_{_\II} =   \sum_{n=0}^\infty \LL_{_\II}^n\,, \qquad \LL_{_\FP} =   \sum_{n=0}^\infty \LL_{_\FP}^n\,,
\\
\label{man-24092016-11} && \LL_\tau^n \equiv  \frac{e}{2n!} \phi_\tau^{a_1\ldots a_n} (\Box_\sm(A)dS + M_n) \phi_\tau^{a_1\ldots a_n}\,,\qquad \tau = \I,\II
\\
\label{man-24092016-14} && \LL_{_\FP}^n \equiv  \frac{e}{n!} \cb^{a_1\ldots a_n} (\Box_\sm(A)dS + M_n) c^{a_1\ldots a_n}\,,
\\
\label{man-24092016-15} && \hspace{1cm} M_n \equiv - \mu_0 - \rho \Bigl( n(n+d-1) + 2d - 4 \Bigr).
\eeq
From \rf{man-24092016-03}, \rf{man-24092016-11},\rf{man-24092016-14}, we see that the partition function of continuous spin field is given by

\beq
\label{man-24092016-16} && \hspace{-1cm} Z = Z_{\FP}/{ Z_{_\I}Z_{_\II} }\,,
\\
\label{man-24092016-17} &&  Z_{_\I} = \ZZ \,, \qquad  Z_{_\II} = \ZZ\,, \qquad Z_{\FP} = \ZZ^2\,,
\\
\label{man-24092016-18} && \ZZ \equiv \prod_{n=0}^\infty \DD_n(M_n) \,, \qquad \DD_n (M_n) \equiv \sqrt{\det{}\!_n(- \Box_\sm(A)dS - M_n)}\,,
\eeq
where in relation \rf{man-24092016-18} the determinant of D'Alembert operator in (A)dS is evaluated on space of traceless rank-$n$ tensor field.  Using \rf{man-24092016-17}, we see that partition function of continuous spin field \rf{man-24092016-16} is indeed equal to 1, $Z  = 1$.
Note that the partition function of continuous spin field turns out to be equal to 1 without the use of any special regularization procedure required for a computation of partition functions in higher-spin gauge field theory (see, e.g., Ref.\cite{Beccaria:2015vaa}). For continuous spin field in (A)dS, we note the same mechanism of cancellation as for higher-spin fields in flat space (see Eq.(2.2) in Ref.\cite{Beccaria:2015vaa}). Namely, using \rf{man-24092016-11}-\rf{man-24092016-15}, we check the cancellation of determinant of the physical spin-$n$ field and ghost determinant of spin-$(n+1)$ field. To this end, we note that partition function $Z$ for Lagrangian $\LL_\tot$ \rf{man-24092016-03} can alternatively be represented as
\be \label{man-26122016-01}
Z = \prod_{n=0}^\infty Z_n, \qquad Z_n =  \frac{\DD_{n-1}(M_{n-1})\DD_{n-1}(M_{n-1})}{\DD_n(M_n)\DD_{n-2}(M_{n-2})}\,, \quad \DD_{-2}(M_{-2}) \equiv 1,  \quad \DD_{-1}(M_{-1}) \equiv 1 \,.
\ee
We now use the following relations:
\be \label{man-26122016-02}
\DD_n(M_n) = \DD_n^\perp (M_n) \DD_{n-1}(M_{n-1}) \,, \qquad
Z_n =  \frac{\DD_{n-1}^\perp(M_{n-1})}{\DD_n^\perp(M_n)}\,,
\ee
where $\DD_n^\perp(M_n)$ takes the form as in \rf{man-24092016-18} with a prescription that the determinant of D'Alembert operator in (A)dS is evaluated on space of traceless and divergence-free rank-$n$ tensor field in $(A)dS_{d+1}$. From $Z_n$ \rf{man-26122016-02}, we see the cancellation of determinant of the physical spin-$n$ field and ghost determinant of spin-$(n+1)$ field in expression for $Z$ in \rf{man-26122016-01}.

\newsection{\large  (Ir)reducible classically unitary continuous spin field}

Lagrangian \rf{man-20092016-04} depends on the two arbitrary real-valued parameters $\mu_0$, $\mu_1$. Our aim in this Section is to find restrictions imposed on these parameters for reducible and irreducible classically unitary systems. We start with our definition of classically unitary (ir)reducible systems.

\noindent \ibf) Lagrangian \rf{man-20092016-04} is formulated in terms of real-valued fields \rf{man-19092016-01} and real-valued gauge transformation parameters \rf{man-21092016-01}. Therefore in order for gauge variation of fields \rf{man-21092016-06} to be real-valued the quantity $F(n)$ \rf{man-20092016-14} should be positive for all $n=0,1,\ldots,\infty$. Let us represent $F(n)$ \rf{man-20092016-14} as

\be \label{man-01102016-01}
F(n) = \mu_1 - \bigl(\mu_0 + \rho (d-3)\bigr) x_n  - \rho x_n^2   \,,\qquad  x_n \equiv n (n + d-2)\,.
\ee
If $F(n)$ \rf{man-01102016-01} is positive for all $n$, then fields \rf{man-19092016-01} will referred to as classically unitary system,
\be \label{man-27092016-01}
F(n)\geq 0 \ \hbox{ for all } \ n=0,1,\ldots, \infty  \hspace{1cm} \hbox{ classically unitary system}.
\ee

\noindent \iibf) If $F(n)$ \rf{man-01102016-01} has no roots,  then fields \rf{man-19092016-01} will be referred to as irreducible dynamical system, while if $F(n)$ \rf{man-01102016-01} has roots, then fields \rf{man-19092016-01} will be referred to as reducible system,
\beq
\label{man-27092016-02} && F(n) \ne 0 \ \hbox{ for all } \ n =  0,1,\ldots, \infty,  \hspace{1.6cm} \hbox{ irreducible system},
\\
\label{man-27092016-03} && F(n_r) = 0 \ \hbox{ for some } \ n_r \in 0,1,\ldots, \infty,  \hspace{1cm} \hbox{ reducible system}.
\eeq
For $F(n)$ in \rf{man-27092016-02}, Lagrangian \rf{man-20092016-04} describes infinite chain of coupling fields \rf{man-19092016-01}, while, for $F(n)$ in \rf{man-27092016-03}, Lagrangian \rf{man-20092016-04} is factorized and describes finite and infinite decoupled chains of fields.

We are interested in (ir)reducible classically unitary systems. Using definitions \rf{man-27092016-01}-\rf{man-27092016-03}, an irreducible classically unitary system is defined by relations
\be
\label{man-27092016-04} F(n) > 0 \ \hbox{ for all } \ n =0,1,\ldots, \infty,  \hspace{1cm} \hbox{ irreducible classically unitary system};
\ee
while reducible classically unitary system  is defined by relations
\beq
&& \hspace{-1.4cm}  F(n_r) = 0  \ \hbox{ for some } n_r \in  0,1,\ldots, \infty\,,
\nonumber\\
\label{man-27092016-05} && \hspace{-1.4cm} F(n) > 0  \ \hbox{ for all } n =0,1,\ldots, \infty \hbox{ and } n \ne n_r \hspace{1cm} \hbox{ reducible classically unitary system}.
\eeq

We now consider (ir)reducible classically unitary systems for flat, AdS, and dS spaces in turn.

\noindent {\bf Flat space, $\rho=0$}. Plugging $\rho=0$ in \rf{man-01102016-01}, we find that restrictions on  $F(n)$ \rf{man-27092016-04}, \rf{man-27092016-05} lead to the following allowed values for the parameters $\mu_0$ and $\mu_1$

\beq
\label{man-27092016-06} &&  \mu_0\leq 0\,, \qquad \mu_1 > 0\,, \qquad  \hbox{ for irreducible classically unitary system},
\\
\label{man-27092016-07} && \mu_0 =0\,, \qquad \mu_1 = 0\,, \qquad  \hbox{ for reducible classically unitary system}.
\eeq
From \rf{man-24092016-11}, we see that cases $\mu_0 = 0$, $\mu_0 > 0$, and $\mu_0 < 0 $ are associated with the respective massless, massive, and tachyonic fields in flat space.
Note that, for the case of \rf{man-27092016-07}, we have $F(n)\equiv 0$ and this case describes a chain of standard massless fields in flat space which consists of every spin just once. Case $\mu_0=0$, $\mu_1>0$ \rf{man-27092016-06} describes massless continuous spin field in flat space.%
\footnote{ For $\mu_0=0$, $\mu_1>0$, one can make sure that our Lagrangian \rf{man-20092016-04} leads to light-cone gauge description of massless continuous spin field discussed in Ref.\cite{Brink:2002zx}. Therefore we think that the case $\mu_0=0$, $\mu_1>0$  is associated with massless continuous spin field in flat space. Also, using the covariant gauge $\Lb\phik=0$ with $\Lb$ as in \rf{man-20092016-06}, we find that our Lagrangian \rf{man-20092016-04} leads to equations of motion $(\Box-\mu_0)\phik =0$, i.e., the case $\mu_0=0$ describes massless field.}
For such field in $R^{3,1}$, gauge invariant Lagrangian was obtained in  Ref.\cite{Schuster:2014hca}. Case $\mu_0 < 0$, $\mu_1>0$ \rf{man-27092016-06} describes tachyonic continuous spin field in flat space. To our knowledge Lagrangian gauge invariant description for massless continuous spin field in $R^{d,1}$, $d>3$, and tachyonic continuous spin field in $R^{d,1}$, $d\geq 3$, has not been discussed in earlier literature. We expect that our continuous spin field with $\mu_0 < 0$, $\mu_1 > 0$ \rf{man-27092016-06} is associated with tachyonic UIR of Poincar\'e algebra.%
\footnote{ Discussion of group theoretical aspects of tachyonic UIR of the Poincar\'e algebra may be found in Ref.\cite{Bekaert:2006py}.}

If we ignore restrictions in \rf{man-27092016-01}, \rf{man-27092016-02}, then restriction \rf{man-27092016-03} leads to some interesting reducible dynamical system. Namely, plugging $\rho=0$ in \rf{man-01102016-01}, we note that equation $F(s)=0$ implies
\be \label{man-28092016-01}
\mu_1 =  s (s +d-2)\mu_0\,.
\ee
Plugging \rf{man-28092016-01} into \rf{man-20092016-14} with $\rho=0$, we find

\be \label{man-28092016-02}
F(N_\upsilon) =  (s-N_\upsilon) (s +d-2 + N_\upsilon) \mu_0\,.
\ee
Using $F(N_\upsilon)$ \rf{man-28092016-02}, we can check that Lagrangian \rf{man-20092016-04} and gauge transformations \rf{man-21092016-06} describe reducible system of gauge fields \rf{man-20092016-01}. Namely, decomposing ket-vector $\phik$ \rf{man-20092016-01} as

\beq
\label{man-28092016-03} \phik & = &  |\phi^{0,s}\rangle  + |\phi^{s+1,\infty}\rangle\,,
\\
\label{man-28092016-05} && |\phi^{M,N}\rangle \equiv  \sum_{n=M}^N \frac{\upsilon^n}{n!\sqrt{n!}} \alpha^{a_1} \ldots \alpha^{a_n} \phi^{a_1\ldots a_n} |0\rangle\,,
\eeq
we verify that Lagrangian \rf{man-20092016-04} with $\rho=0$ and $F(N_\upsilon)$ \rf{man-28092016-02}  is factorized as
\be
\label{man-28092016-06}  \LL = \LL^{0,s} + \LL^{s+1,\infty}\,, \qquad \LL^{0,s} \equiv \half \langle \phi^{0,s}|E|\phi^{0,s}\rangle\,,  \qquad \LL^{s+1,\infty} \equiv
\half \langle \phi^{s+1,\infty}|E|\phi^{s+1,\infty}\rangle\,,
\ee
where $E$ is given in \rf{man-20092016-05}. Also it easy to check that gauge transformations \rf{man-21092016-06} with $\rho=0$ and $F(N_\upsilon)$ as in \rf{man-28092016-02} are also factorized.
Namely, decomposing ket-vector $\xik$ \rf{man-21092016-03} as

\beq
\label{man-28092016-08} && \xik =  |\xi^{0,s-1}\rangle  + |\xi^{s,\infty}\rangle \,,
\\
\label{man-28092016-10} && \hspace{1cm} |\xi^{M,N}\rangle \equiv \sum_{n=M}^N \frac{\upsilon^{n+1}}{n!\sqrt{(n+1)!}} \alpha^{a_1} \ldots \alpha^{a_n} \xi^{a_1\ldots a_n} |0\rangle\,,
\eeq
we check that gauge transformations \rf{man-21092016-06} with $\rho=0$ and $F(N_\upsilon)$  \rf{man-28092016-02} are also factorized,
\be \label{man-28092016-11}
\delta |\phi^{0,s}\rangle  = G|\xi^{0,s-1}\rangle\,, \qquad \delta |\phi^{s+1,\infty}\rangle  = G |\xi^{s,\infty}\rangle\,.
\ee
Thus we see that if $\mu_0$ and $\mu_1$ are related as in \rf{man-28092016-01} then our Lagrangian \rf{man-20092016-04} describes two decoupling fields $|\phi^{0,s}\rangle$ and  $|\phi^{s+1,\infty}\rangle$ \rf{man-28092016-05}. If $\mu_0> 0$, then we use $\mu_0=m^2$ and note that $|\phi^{0,s}\rangle$ describes a classically unitary spin-$s$ and mass-$m$ massive field. For $\mu_0>0$, we have $F(n) < 0$ when $n=s+1,s+2,\ldots,\infty$ and therefore $|\phi^{s+1,\infty}\rangle$ describes classically non-unitary system. In contrary, if $\mu_0 <  0$, then $|\phi^{0,s}\rangle$ describes a classically non-unitary spin-$s$ massive field. For $\mu_0<0$, we have $F(n) > 0$ when $n=s+1,s+2,\ldots,\infty$ and therefore $|\phi^{s+1,\infty}\rangle$ describes classically unitary system. We refer to such $|\phi^{s+1,\infty}\rangle$ as spin-$(s+1)$ infinite-component field.%
\footnote{ In $R^{d,1}$, such field seems to be related to the tachyonic representation with a discrete series for its little algebra $so(1,d-1)$. We thank the referee for pointing this out to us.
}

\bigskip
\noindent {\bf (A)dS space, $\rho \ne 0$}. We now study reducible and irreducible classically unitary systems for (A)dS space, i.e., we study solution of Eqs.\rf{man-27092016-04},\rf{man-27092016-05}. Our study of \rf{man-27092016-04} \rf{man-27092016-05} is summarized as follows

\bigskip
\noindent {\bf Statement 1}. {\it For dS space, equations \rf{man-27092016-04}, \rf{man-27092016-05} do not have solutions}. Absence of solution of equation \rf{man-27092016-04} implies that continuous spin field in dS space is not realized as irreducible and classically unitary  system, while absence of solution of equation \rf{man-27092016-05} implies that continuous spin field in dS space is not realized as reducible and classically unitary systems.

\bigskip
\noindent {\bf Statement 2}. {For AdS space, equations \rf{man-27092016-04} have solutions which we classify as Type I,II, and III solutions},%
\footnote{ Representations of the $so(d,2)$ algebra which are associated with our classically unitary irreducible systems of gauge fields in $AdS_{d+1}$ are still to be identified. Note that all previously known examples of classically unitary irreducible systems of {\it gauge} fields described by Lagrangian with positive sign in d'Alembertian operators are related to {\it unitary} irreps of the $so(d,2)$ algebra. Our Lagrangian has  positive sign in d'Alembertian operators. Therefore we believe that our classically unitary irreducible systems are also related to {\it unitary} irreps of the $so(d,2)$ algebra.}

\noindent {\it Type I solutions for AdS:}

\be \label{man-29092016-01}
\mu_0 - |\rho| (d-3) \leq  0\,, \qquad \mu_1 > 0\,.
\ee
\noindent {\it Type II solutions for AdS:}

\beq
\label{man-29092016-02} && \mu_0 =  |\rho| (d-3)  + 2|\rho|\lambda_0 (\lambda_0 + d-2)\,,
\nonumber\\
&& \mu_1 >   |\rho| \lambda_0^2 (\lambda_0 + d-2)^2\,, \qquad \lambda_0 = 1,2,\ldots, \infty\,.
\eeq
\noindent {\it Type III solutions for AdS:}

\beq
\label{man-29092016-03} && \mu_0 =  |\rho| (d-3)  + 2|\rho|\lambda (\lambda + d-2)\,,
\\
\label{man-29092016-04} && \mu_1 >   |\rho| \lambda^2 (\lambda + d-2)^2   -|\rho| \epsilon^2 (\epsilon + 2\lambda_0 + d-2)^2  \hspace{1.9cm} \hbox{ for } \ 0 < \epsilon < \epsilon_r\,,
\\
\label{man-29092016-05} && \mu_1 >   |\rho| \lambda^2 (\lambda + d-2)^2  -|\rho| (1-\epsilon)^2 (\epsilon + 2\lambda_0 + d-1)^2\,,\hspace{0.6cm} \hbox{ for } \ \epsilon_r < \epsilon < 1\,,\qquad
\\
\label{man-29092016-06} && \mu_1 >   |\rho| \lambda^2 (\lambda + d-2)^2 - \frac{1}{4}|\rho|(2\lambda_0 + d-1)^2\,,  \hspace{2.4cm} \hbox{ for } \ \epsilon = \epsilon_r\,,
\\
\label{man-29092016-07} && \lambda = \lambda_0+\epsilon\,, \qquad 0 < \epsilon < 1\,, \qquad \lambda_0=0,1,\ldots,\infty \,,
\\
\label{man-29092016-08} && \epsilon_r \equiv  \bigl( \sqrt{ (2\lambda_0 + d-1)^2 + 1} - 2\lambda_0 - d + 2\bigr)/2.
\eeq
We note that type I solutions are obtained by considering $\mu_0 \leq |\rho| (d-3)$, while type II and III solutions are obtained by considering $\mu_0 > |\rho| (d-3)$. Note also that type II solutions are labelled by integer $\lambda_0$ \rf{man-29092016-02}, while type III solutions are labelled by $\epsilon$,  $0<\epsilon<1$, and integer $\lambda_0$ \rf{man-29092016-07}.

\bigskip
\noindent {\bf Statement 3}. {\it For AdS space, equations \rf{man-27092016-05} have solutions  given by}
\beq
\label{man-30092016-01} && \mu_1 =  s (s +d-2) \Bigl(\mu_0 - |\rho| (s + 1)(s + d-3) \Bigr) \,,
\\
\label{man-30092016-02} && 2 |\rho| s(s+d-3) < \mu_0 < 2 |\rho| (s+1)(s+d-2) \,.
\eeq
Lagrangian \rf{man-20092016-04} with  $\mu_1$ given in \rf{man-30092016-01} describes reducible system of continuous spin field \rf{man-20092016-01}. Namely, decomposing $\phik$ \rf{man-20092016-01} as
\be \label{man-30092016-03}
\phik = |\phi^{0,s}\rangle  + |\phi^{s+1,\infty}\rangle \,,
\ee
where $|\phi^{M,N}\rangle$ is defined in \rf{man-28092016-05}, we check that Lagrangian \rf{man-20092016-04} with $\mu_1$ \rf{man-30092016-01} is factorized as

\be \label{man-30092016-04}
\LL = \LL^{0,s} + \LL^{s+1,\infty}\,, \qquad \LL^{0,s} \equiv \frac{e}{2} \langle \phi^{0,s}|E|\phi^{0,s}\rangle\,,  \qquad \LL^{s+1,\infty} \equiv
\frac{e}{2} \langle \phi^{s+1,\infty}|E|\phi^{s+1,\infty}\rangle\,,
\ee
where $E$ is given in \rf{man-20092016-05}. Gauge transformations \rf{man-21092016-06} with $\mu_1$ given in \rf{man-30092016-01} are also factorized. Namely, decomposing ket-vector $\xik$ \rf{man-21092016-03} as
\be
\xik =  |\xi^{0,s-1}\rangle  + |\xi^{s,\infty}\rangle \,,
\ee
we verify that gauge transformations \rf{man-21092016-06} with $\mu_0$, $\mu_1$ as in \rf{man-30092016-01},\rf{man-30092016-02} are also factorized,
\be
\delta |\phi^{0,s}\rangle  = G|\xi^{0,s-1}\rangle\,, \qquad \delta |\phi^{s+1,\infty}\rangle  = G |\xi^{s,\infty}\rangle\,.
\ee

The Statements are proved by noticing that $F(n)$ \rf{man-01102016-01} has at most two roots, i.e., we have three cases: 1) $F(n)$ has no roots; 2) $F(n)$ has one root; 3) $F(n)$ has two roots; We analyse these cases in turn.

\noindent \ibf) Using \rf{man-01102016-01}, we note that equations \rf{man-27092016-04} can alternatively be represented as
\be \label{man-29092016-09}
\mu_1 >  \max_{n=0,1,\ldots,\infty} \Bigl( \bigl(\mu_0 + \rho (d-3)\bigr) x_n  +    \rho x_n^2 \Bigr)\,.
\ee
We now see that, for dS space ($\rho > 0$), equation \rf{man-29092016-09} has no solution, while, for AdS space ($\rho = -|\rho|$) with $\mu_0 \leq |\rho| (d-3)$, equation \rf{man-29092016-09} implies $\mu_1>0$.
Analysis of the case $\mu_0> |\rho|(d-3)$ leads to Type II and III solutions.

\noindent \iibf) Now we analyse the case when $F(n)$ has one root $n_r=s$. Then Eqs.\rf{man-27092016-05} amount to

\be \label{man-01102016-02}
F(s) = 0\,, \qquad  F(n) > 0 \quad \hbox{ for } \quad n =0,1,\ldots, s-1, s+1,s+2,\ldots, \infty\,.
\ee
It is easy to check that Eqs.\rf{man-01102016-02} lead to the following restrictions on $\mu_0$, $\mu_1$
\beq
\label{man-01102016-03} &&    \mu_1 =  s (s +d-2) \Bigl(\mu_0 + \rho (s + 1)(s + d-3) \Bigr) \,,
\\
&& - 2 \rho  s(s+d-3) < \mu_0 <   - 2\rho (s+1)(s+d-2) \,, \hspace{1.5cm} \hbox{ for AdS}\,,
\nonumber\\[-10pt]
\label{man-01102016-04} &&
\\[-10pt]
&& - \rho  (s+1) (s+d-3) <  \mu_0 <   -\rho \infty \,, \hspace{3.7cm} \hbox{ for dS}\,.
\nonumber
\eeq
Namely, $\mu_1$ \rf{man-01102016-03} is obtained from equation $F(s)=0$. Left inequality in \rf{man-01102016-04} is obtained by requiring $F(n)>0$ for $n=0,1,\ldots,s-1$,
while right inequality in \rf{man-01102016-04} is obtained by requiring $F(n)>0$ for $n=s+1,\ldots, \infty$. Plugging $\mu_1$ \rf{man-01102016-03} in \rf{man-20092016-14} and using the interrelation between $\mu_0$ and a standard mass parameter $m^2$,
\be \label{man-01102016-04-a1}
\mu_0 + 2\rho s (s + d - 3) = m^2\,,
\ee
we cast $F(N_\upsilon)$ \rf{man-20092016-14} into the form
\be  \label{man-01102016-04-a2}
F(N_\upsilon) =  (s-N_\upsilon) (s +d-2 + N_\upsilon)\Bigl(m^2 - \rho (s-1-N_\upsilon)(s + d-3 + N_\upsilon) \Bigr)\,.
\ee
Lagrangian \rf{man-20092016-04} with $F(N_\upsilon)$ \rf{man-01102016-04-a2} describe reducible system of continuous spin field \rf{man-20092016-01}. Namely, decomposing $\phik$ \rf{man-20092016-01} as in \rf{man-30092016-03}
we verify that Lagrangian \rf{man-20092016-04} is factorized as in \rf{man-30092016-04}.
For dS space, relation $\rho=|\rho|$ implies that inequalities \rf{man-01102016-04} are inconsistent. Thus, for dS space, Eqs.\rf{man-27092016-05} with one root $n_r=s$ do not have solutions.
For $AdS$, using $\rho=-|\rho|$, we see that \rf{man-01102016-03},\rf{man-01102016-04} lead to \rf{man-30092016-01},\rf{man-30092016-02}. Note that, using \rf{man-01102016-04-a1}, we can represent \rf{man-01102016-04} as
\be
0 < m^2 <  - 2\rho (2s + d-2) \qquad \hbox{ for AdS}\,.
\ee

\noindent \iiibf) Finally we analyse Eqs.\rf{man-27092016-05} for the case when $F(n)$ \rf{man-01102016-01} has two roots $n_r=s,S$, $s \leq S$,
\be  \label{man-01102016-04-a3}
F(s) = 0 \,, \qquad F(S) = 0\,, \qquad s \leq S\,.
\ee
Using $F(n)$ \rf{man-01102016-01},  it is easy to check that solution to equations \rf{man-01102016-04-a3} is given by
\be
\label{man-01102016-05}  \mu_0 = - \rho s(s+d-2) - \rho (S+1)(S + d-3)\,, \qquad \mu_1 = - \rho s(s+d-2) S(S+d-2) \,.
\ee
Plugging \rf{man-01102016-05} into \rf{man-20092016-14}, we get
\be \label{man-01102016-07}
F(N_\upsilon) =  - \rho (s-N_\upsilon) (s +d-2 + N_\upsilon)(S-N_\upsilon) (S +d-2 + N_\upsilon)\,.
\ee
Lagrangian \rf{man-20092016-04} and gauge transformations \rf{man-21092016-06} with  $\mu_0$, $\mu_1$ \rf{man-30092016-01},\rf{man-30092016-02} and $F(N_\upsilon)$ \rf{man-01102016-07} describe reducible system of continuous spin field \rf{man-20092016-01}. Namely, decomposing $\phik$ \rf{man-20092016-01} as
\be \label{man-01102016-08}
\phik = |\phi^{0,s}\rangle  + |\phi^{s+1,S}\rangle + |\phi^{S+1,\infty}\rangle \,,
\ee
where $|\phi^{M,N}\rangle$ is defined in \rf{man-28092016-05}, we verify that Lagrangian \rf{man-20092016-04} is factorized as
\be \label{man-01102016-09}
\LL = \LL^{0,s} + \LL^{s+1,S} + \LL^{S+1,\infty}\,, \qquad \LL^{M,N} \equiv \frac{e}{2} \langle \phi^{M,N}|E|\phi^{M,N}\rangle\,,
\ee
where $E$ is given in \rf{man-20092016-05}. In \rf{man-01102016-08}, we assume that if $s=S$, then  $|\phi^{s+1,s}\rangle\equiv 0$.  From \rf{man-01102016-07},\rf{man-01102016-09}, we learn that $|\phi^{0,s}\rangle$ \rf{man-01102016-08} describes classically unitary (non-unitary) massive spin-$s$ field in AdS space (dS space), the $|\phi^{s+1,S}\rangle$ \rf{man-01102016-08} describes classically non-unitary partial-massless spin-$S$ field in (A)dS, while the $|\phi^{S+1,\infty}\rangle$ \rf{man-01102016-08} describes classically unitary (non-unitary) spin-$(S+1)$ infinite-component field in AdS space (dS space). Mass parameters of $|\phi^{0,s}\rangle$ and $|\phi^{s+1,S}\rangle$ are given by
\beq
\label{man-02102016-01}  && m^2 = -\rho (S+1-s)(S+s + d-3)\,, \hspace{2.7cm} \hbox{ for } |\phi^{0,s}\rangle\,,
\\
\label{man-02102016-02} && m_k^2 \equiv  \rho k (2S+d-4-k)\,, \qquad k\equiv S-s-1\,, \hspace{1cm} \hbox{ for } |\phi^{s+1,S}\rangle\,.
\eeq
Relation \rf{man-02102016-02} tells us that $|\phi^{s+1,S}\rangle$ is a depth-$k$ partial-massless field. For $S=s+1$, this field turns out to be spin-$(s+1)$ massless field.

\medskip

To summarize, in this paper, we developed gauge invariant Lagrangian formulation for continuous
spin field in (A)dS and applied our result for a computation of partition function.
In this paper, we used metric-like Lagrangian formulation of gauge fields.%
\footnote{ We expect that, at level of light-cone gauge equations of motion and unfolding equations of motion,  a description of continuous spin AdS field appeared in the respective Ref.\cite{Metsaev:1999ui} and Ref.\cite{Ponomarev:2010st}. Namely, we expect that, in Sec.5 in Ref.\cite{Metsaev:1999ui}, continuous spin is realized for the arbitrary parameters $\lambda^{+-}$ and $\lambda$, while, in Sec. 2 in Ref.\cite{Ponomarev:2010st}, continuous spin is realized for the arbitrary parameters $l_0$ and $k_0$. We thank E.D.Skvortsov and M.A.Vasiliev for pointing out Ref.\cite{Ponomarev:2010st} to us.
}
In the literature, there are
many interesting approaches to Lagrangian formulation of gauge fields.
We mention frame-like formulation and BRST approach (see, e.g., Refs.\cite{Alkalaev:2005kw,Buchbinder:2011xw}). It will be interesting to study continuous spin field in the framework of such formulations and establish their connection with a vector-superspace formulation in Refs.\cite{Schuster:2014hca,Rivelles:2014fsa}. We note also that use of extended
hamiltonian approach could be helpful for better understanding of physical d.o.f for
continuous spin field. Recent discussion of extended hamiltonian approach may be found in Refs.\cite{Metsaev:2011iz,Campoleoni:2016uwr}. Applications of various methods, which were developed for analysis of interaction vertices of gauge fields in Refs.\cite{Vasilev:2011xf}-\cite{Grigoriev:2016bzl}, to the study of interaction vertices of continuous spin field could of some interest. Light-cone gauge methods in Refs.\cite{Bengtsson:1983pd}-\cite{Metsaev:2007rn} can also be helpful for this purpose.
Recent developments in light-cone approach to field dynamics may be found in Refs.\cite{Bengtsson:2014qza}.

\bigskip
{\bf Acknowledgments}. We thank X. Bekaert for informal seminar on continuous spin field given at Lebedev Institute in June 2016. The seminar triggered our interest in continuous spin field. This work was supported by the RFBR Grant No.14-02-01171.

\setcounter{section}{0} \setcounter{subsection}{0}
\appendix{ \large  Notation }

Vector indices of the $so(d,1)$ algebra  take the values $a,b,c=0,1,\ldots
,d$. We use mostly positive flat metric tensor
$\eta^{ab}$. To simplify our expressions we drop $\eta_{ab}$ in the scalar products.
We use the creation operators $\alpha^a$, $\upsilon$ and the
respective annihilation operators $\alphab^a$, $\upsilonb$,

\be
[\bar{\alpha}^a,\alpha^b]=\eta^{ab}\,,  \quad [\upsilonb,\upsilon]=1\,, \quad
\bar\alpha^a |0\rangle = 0\,,\quad  \upsilonb |0\rangle = 0\,, \qquad \alpha^{a\dagger} = \alphab^a, \qquad \upsilon^{\dagger} = \upsilonb.
\ee
These operators are referred to as oscillators in this paper. The oscillators
$\alpha^a$, $\bar\alpha^a$ and $\upsilon$, $\upsilonb$, transform in the respective vector and scalar representations of the $so(d,1)$ algebra.
Realization of covariant derivative $D^a$ on space of ket-vector $\phik$ \rf{man-20092016-01}  is given by $D^a = \eta^{ab}D_b$,

\be \label{man-24092016-21}
D_a \equiv e_a^\mun D_\mun\,,  \qquad D_\mun \equiv
\partial_\mun +\frac{1}{2}\omega_\mun^{ab} M^{ab}\,, \qquad \partial_\mun = \partial/\partial x^\mun\,, \qquad M^{ab} \equiv \alpha^a
\alphab^b - \alpha^b \alphab^a\,,
\ee
where base manifold index takes values $\mun = 0,1,\ldots, d$.
In \rf{man-24092016-21}, $D_\mun$ stands for the Lorentz covariant derivative, while $e_a^\mun$ is inverse vielbein of $AdS_{d+1}$ space. Also note that $M^{ab}$ is a spin
operator of the Lorentz algebra $so(d,1)$, while $\omega_\mun^{ab}$ is the Lorentz connection of $AdS_{d+1}$ space. $AdS_{d+1}$ space contravariant tensor field, $\phi^{\mun_1\ldots \mun_n}$, is related with field carrying the flat indices, $\phi^{a_1\ldots
a_n}$, in a standard way $\phi^{a_1\ldots a_n} \equiv e_{\mun_1}^{a_1}\ldots
e_{\mun_n}^{a_n} \phi^{\mun_1\ldots \mun_n}$. We use the conventions

\beq
\label{man-24092016-21-a1} && \alpha^2 \equiv \alpha^a\alpha^a\,, \hspace{1cm} \alphab^2 \equiv \alphab^a\alphab^a\,,  \hspace{1cm} N_\alpha \equiv \alpha^a \alphab^a\,,  \hspace{1cm} N_\upsilon \equiv \upsilon \upsilonb\,,
\\
\label{man-24092016-21-a2}&& \alpha D \equiv \alpha^a D^a\,, \hspace{1cm} \alphab D \equiv \alphab^a D^a\,, \hspace{0.5cm} \Box_\sm(A)dS \equiv D^a D^a + e^{a\mun}\omega_\mun^{ab}D^b\,,
\\
\label{man-24092016-21-a3} && \Pi^\smponetwo \equiv 1 - \alpha^2 \frac{1}{2(2N_\alpha + d + 1)}\alphab^2\,.
\eeq


\vspace{1cm}

\small
\begin{thebibliography}{30}

\parskip=-1pt


\bibitem{Bargmann:1948ck}
  V.~Bargmann and E.~P.~Wigner,
  Proc.\ Nat.\ Acad.\ Sci.\  {\bf 34}, 211 (1948).


\bibitem{Brink:2002zx}
  L.~Brink, A.~M.~Khan, P.~Ramond and X.~z.~Xiong,
  J.\ Math.\ Phys.\  {\bf 43}, 6279 (2002)
  [hep-th/0205145].



\bibitem{Bekaert:2005in}
  X.~Bekaert and J.~Mourad,
  JHEP {\bf 0601}, 115 (2006)
  [hep-th/0509092].




\bibitem{Vasiliev:1990en}
  M.~A.~Vasiliev,
  Phys.\ Lett.\ B {\bf 243}, 378 (1990).
%
\\
%
  M.~A.~Vasiliev,
  Phys.\ Lett.\  B {\bf 567}, 139 (2003)
  [arXiv:hep-th/0304049].



\bibitem{Savvidy:2003fx}
  G.~K.~Savvidy,
  Int.\ J.\ Mod.\ Phys.\ A {\bf 19}, 3171 (2004)
  [hep-th/0310085].
%
\\
%
  J.~Mourad,
  ``Continuous spin particles from a string theory,''
  hep-th/0504118.




\bibitem{Schuster:2014hca}
  P.~Schuster and N.~Toro,
  Phys.\ Rev.\ D {\bf 91}, 025023 (2015)
  [arXiv:1404.0675 [hep-th]].


\bibitem{Najafizadeh:2015uxa}
  X.Bekaert, M.Najafizadeh, M.R.Setare,
  Phys.\ Lett.\ B {\bf 760}, 320 (2016)
  [arXiv:1506.00973 [hep-th]].



\bibitem{Metsaev:2014vda}
  R.~R.~Metsaev,
  Theor.\ Math.\ Phys.\  {\bf 181}, no. 3, 1548 (2014)
  [arXiv:1407.2601 [hep-th]].



\bibitem{Bengtsson:2013vra}
  A.~K.~H.~Bengtsson,
  JHEP {\bf 1310}, 108 (2013)
  [arXiv:1303.3799 [hep-th]].




\bibitem{Metsaev:2008fs}
  R.~R.~Metsaev,
  Phys.\ Rev.\  D {\bf 78}, 106010 (2008)
  [arXiv:0805.3472 [hep-th]].



\bibitem{Metsaev:2009hp}
  R.~R.~Metsaev,
  Phys.\ Lett.\ B {\bf 682}, 455 (2010)
  [arXiv:0907.2207 [hep-th]].



\bibitem{Metsaev:2014iwa}
  R.~R.~Metsaev,
  Nucl.\ Phys.\ B {\bf 885}, 734 (2014)
  [arXiv:1404.3712 [hep-th]].



\bibitem{Zinoviev:2001dt}
Yu.~M.~Zinoviev, ``On massive high spin particles in (A)dS,''
arXiv:hep-th/0108192.


\bibitem{Francia:2007qt}
  D.~Francia, J.~Mourad and A.~Sagnotti,
  Nucl.\ Phys.\ B {\bf 773}, 203 (2007)
  [hep-th/0701163].
%
\\
%
  S.~Guttenberg and G.~Savvidy,
  SIGMA {\bf 4}, 061 (2008)
  [arXiv:0804.0522 [hep-th]].
%
\\
%
  A.~Fotopoulos and M.~Tsulaia,
  JHEP {\bf 0910}, 050 (2009)
  [arXiv:0907.4061 [hep-th]].
%
\\
%
  R.~Manvelyan, K.~Mkrtchyan and W.~Ruhl,
  Nucl.\ Phys.\  B {\bf 803}, 405 (2008)
  [arXiv:0804.1211 [hep-th]].
%
\\
%
  D.~Ponomarev and A.~A.~Tseytlin,
  JHEP {\bf 1605}, 184 (2016)
  [arXiv:1603.06273 [hep-th]].


\bibitem{Beccaria:2015vaa}
  M.~Beccaria and A.~A.~Tseytlin,
  J.\ Phys.\ A {\bf 48}, no. 27, 275401 (2015)
  [arXiv:1503.08143 [hep-th]].



\bibitem{Bekaert:2006py}
  X.~Bekaert and N.~Boulanger,
  ``The Unitary representations of the Poincare group in any spacetime dimension,''
  hep-th/0611263.




\bibitem{Metsaev:1999ui}
  R.~R.~Metsaev,
  Nucl.\ Phys.\ B {\bf 563}, 295 (1999)
  [hep-th/9906217].


\bibitem{Ponomarev:2010st}
  D.~S.~Ponomarev and M.~A.~Vasiliev,
  Nucl.\ Phys.\ B {\bf 839}, 466 (2010)
  [arXiv:1001.0062 [hep-th]].




\bibitem{Alkalaev:2005kw}
  K.~B.~Alkalaev, O.~V.~Shaynkman and M.~A.~Vasiliev,
  JHEP {\bf 0508}, 069 (2005)
  [hep-th/0501108].
%
\\
%
E.~D.~Skvortsov,
Nucl.\ Phys.\  B {\bf 808}, 569 (2009) [arXiv:0807.0903 [hep-th]];



\bibitem{Buchbinder:2011xw}
  I.~L.~Buchbinder and A.~Reshetnyak,
  Nucl.\ Phys.\ B {\bf 862}, 270 (2012)
  [arXiv:1110.5044 [hep-th]].
%
\\
%
  I.L.Buchbinder, V.A.Krykhtin, A.A.Reshetnyak,
  Nucl.Phys.B {\bf 787}, 211 (2007)
  [hep-th/0703049].
%
\\
%
  I.~L.~Buchbinder, V.~A.~Krykhtin and P.~M.~Lavrov,
  Nucl.\ Phys.\  B {\bf 762}, 344 (2007)
  hep-th/0608005
%
\\
%
  A.~Fotopoulos and M.~Tsulaia,
  Int.\ J.\ Mod.\ Phys.\ A {\bf 24}, 1 (2009)
  [arXiv:0805.1346 [hep-th]].


\bibitem{Rivelles:2014fsa}
  V.~O.~Rivelles,
  Phys.\ Rev.\ D {\bf 91}, no. 12, 125035 (2015)
  [arXiv:1408.3576 [hep-th]].


\bibitem{Metsaev:2011iz}
  R.~R.~Metsaev,
  J.\ Phys.\ A {\bf 46}, 214021 (2013)
  [arXiv:1112.0976 [hep-th]].

\bibitem{Campoleoni:2016uwr}
  A.~Campoleoni, M.~Henneaux, S.~Hörtner and A.~Leonard,
  JHEP {\bf 1610}, 146 (2016)
  [arXiv:1608.04663].
%
\\
%
  M.~Henneaux, S.~Hörtner and A.~Leonard,
  Phys.\ Rev.\ D {\bf 94}, no. 10, 105027 (2016)
  [arXiv:1609.04461].


\bibitem{Vasilev:2011xf}
  M.~A.~Vasiliev,
  Nucl.\ Phys.\ B {\bf 862}, 341 (2012)
  [arXiv:1108.5921 [hep-th]].



\bibitem{Joung:2011ww}
  E.~Joung and M.~Taronna,
  Nucl.\ Phys.\ B {\bf 861}, 145 (2012)
  [arXiv:1110.5918 [hep-th]].
%
\\
%
  E.~Joung, W.~Li and M.~Taronna,
  Phys.\ Rev.\ Lett.\  {\bf 113}, 091101 (2014)
  [arXiv:1406.2335 [hep-th]].

\bibitem{Fotopoulos:2010ay}
  A.~Fotopoulos and M.~Tsulaia,
  JHEP {\bf 1011}, 086 (2010)
  [arXiv:1009.0727 [hep-th]].

\bibitem{Manvelyan:2010je}
  R.~Manvelyan, K.~Mkrtchyan and W.~Ruehl,
  Phys.\ Lett.\ B {\bf 696}, 410 (2011)
  [arXiv:1009.1054 [hep-th]].


\bibitem{Metsaev:2012uy}
  R.~R.~Metsaev,
  Phys.\ Lett.\ B {\bf 720}, 237 (2013)
  [arXiv:1205.3131 [hep-th]].

\bibitem{Boulanger:2013zza}
  N.Boulanger, D.Ponomarev, E.D.Skvortsov and M.Taronna,
  Int.\ J.\ Mod.\ Phys.\ A {\bf 28}, 1350162 (2013)


\bibitem{Grigoriev:2016bzl}
  M.~Grigoriev and A.~A.~Tseytlin,
  arXiv:1609.09381 [hep-th].
%
\\
%
  M.~Grigoriev,
  ``Presymplectic structures and intrinsic Lagrangians,''
  arXiv:1606.07532 [hep-th].



\bibitem{Bengtsson:1983pd}
  A.~K.~H.~Bengtsson, I.~Bengtsson and L.~Brink,
  Nucl.\ Phys.\ B {\bf 227}, 31 (1983).



\bibitem{Metsaev:1991mt}
  R.~R.~Metsaev,
  Mod.\ Phys.\ Lett.\ A {\bf 6}, 359 (1991).

\bibitem{Metsaev:1991nb}
  R.~R.~Metsaev,
  Mod.\ Phys.\ Lett.\ A {\bf 6}, 2411 (1991).


\bibitem{Metsaev:2007rn}
  R.~R.~Metsaev,
  Nucl.\ Phys.\ B {\bf 859}, 13 (2012)
  [arXiv:0712.3526 [hep-th]].



\bibitem{Bengtsson:2014qza}
  A.~K.~H.~Bengtsson,
  JHEP {\bf 1409}, 105 (2014)
  [arXiv:1403.7345 [hep-th]].
%
\\
%
  E.~Conde and A.~Marzolla,
  JHEP {\bf 1609}, 041 (2016)
  [arXiv:1601.08113 [hep-th]].
%
\\
%
  S.~Ananth, L.~Brink and M.~Mali,
  JHEP {\bf 1508}, 153 (2015)
  [arXiv:1507.01068 [hep-th]].
%
\\
%
  E.~Conde, E.~Joung and K.~Mkrtchyan,
  JHEP {\bf 1608}, 040 (2016)
  [arXiv:1605.07402 [hep-th]].
%
\\
%
  C.~Sleight and M.~Taronna,
  arXiv:1609.00991 [hep-th].
%
\\
%
  D.~Ponomarev and E.~D.~Skvortsov,
  arXiv:1609.04655 [hep-th].






\end{thebibliography}
\end{document}